\documentstyle[11pt,epsfig]{article}
\newcommand \be{\begin{equation}}
\newcommand \ba{\begin{eqnarray}}
\newcommand \ee{\end{equation}}
\newcommand \ea{\end{eqnarray}}
\newcommand{\lp}{\left(}
\newcommand{\rp}{\right)}

\begin{document}

\title{Origin of Crashes in 3 US stock markets: Shocks and  Bubbles}

\thispagestyle{empty}

\author{Anders Johansen \\
Teglg\aa rdsvej 119 \\
3050 Humleb\ae k, Denmark \\
email: anders-johansen@get2net.dk} 

\date{\today}
\maketitle

\abstract{This paper presents an exclusive classification of the largest 
crashes in Dow Jones Industrial Average (DJIA), SP500 and NASDAQ in the past 
century. Crashes are objectively defined as the top-rank filtered drawdowns 
(loss from the last local maximum to the next local minimum disregarding 
noise fluctuations), where the size of the filter is determined by the 
historical volatility of the index. It is shown that {\it all} crashes can 
be linked to either an external shock, {\it e.g.}, outbreak of war, {\it or} 
a log-periodic power law (LPPL) bubble with an empirically well-defined 
complex value of the exponent. Conversely, with one sole exception {\it all} 
previously identified  LPPL bubbles are followed by a top-rank drawdown. 
As a consequence, the analysis presented suggest a one-to-one correspondence 
between market crashes defined as top-rank filtered drawdowns on one hand 
and surprising news and LPPL bubbles on the other. We attribute this 
correspondence to the Efficient Market Hypothesis effective on two quite 
different time scales depending on whether the market instability the crash
represent is internally or externally generated.} 

\thispagestyle{empty}
\pagenumbering{arabic}
\newpage
\setcounter{page}{1}

\section{Addressing the Problem} 

The characterization of large negative moves on the stock market is by 
definition of profound importance to risk management of investment 
portfolios. An objective non-arbitrary definition of ``large'' or similarly
``a crash'' has yet to be agreed upon by the financial community. Here, such
a definition of {\it a crash} is presented in terms of $\epsilon$-drawdowns, 
{\it i.e.}, noise filtered drawdowns where the size of the filter $\epsilon$ 
is determined by the historical volatility of the index. An $\epsilon$-drawdown
classifies as a crash {\it if} its rank $i\sim 1$, {\it i.e.}, the 
special name ``crash'' is reserved for the ``few'' largest negative price 
drops. Ideally, one would like to ``prove'' that these largest negative price 
drops are outliers to the bulk of the distribution of price drops. As for now,
a large body of evidence has accumulated which suggest that it this is most
likely the case \cite{outl1,JS2000,outl2,SJ2001,baliproc,JS2002,Lillo} but not 
more. This problem will be further addressed in a future publication and this 
paper only provides for an ``eye-balling'' analysis by the reader with figures 
1-3. The reader is then left to judge for himself but should note that for the 
largest filter used (see figure legend), {\it all} $\epsilon$-drawdowns for all 
3 indices are parametrized by the fit with the function function (\ref{stretched}).

Specifically, for the DJIA the four largest $\epsilon$-drawdowns are 
characterized as crashes, for the SP500 the five largest and for the NASDAQ 
the six largest. Naturally, this does not exclude, {\it e.g.}, that the fifth 
largest $\epsilon$-drawdown in the DJIA may also be characterized as a crash, 
but this is the interpretation of ``few'' put forward here. Returning to the 
other two indices, for the SP500 an additional $\epsilon$-drawdown has been 
included due to the higher historical volatility of the SP500 compared to that 
of the DJIA. For the same reason, the six largest $\epsilon$-drawdowns are 
considered for the NASDAQ.  The time series extend from 01/01 1900 to 17/07 2000 
for the DJIA, from 29/11 1940 to 17/07 2000 for SP500 and from 05/02 1971 to 
17/07 2000 for the NASDAQ.

An $\epsilon$-drawdown is defined as a persistent decrease in the 
price over consecutive days from a local maximum to the next local minimum 
{\it ignoring} price increases in between the two of relative size less than 
$\epsilon$. Since they are constructed from runs of the same sign variations,
$\epsilon$-drawdowns embody a rather subtle dependence which captures the way 
successive drops can influence each other. This ``persistence'' is not measured
by the distribution of returns because, by its very definition, it forgets 
about the relative positions of the returns\footnote{Realizing this allows one
to construct synthetic price data for statistical tests with the same return 
distribution by a reshuffling of the returns\protect\cite{JS2000}}. {\it A priori}, 
the optimal choice of filter should be bounded by $0 < \epsilon \le \sigma$, 
where $\sigma$ is the historical volatility, but as the noise distribution for
financial data is unknown and presumably not stationary over small time scales
a more qualified choice seems difficult to make.

For each of the 12 crashes considered, a causal explanation in terms of
\begin{itemize}
\item  A surprising new piece of information as the Efficient Market 
Hypothesis (EMH) demands. We refer to this as a ``shock''.

\item A prior LPPL ``bubble'', as defined below, to be deflated by the crash 
reminiscent of Adam Smith's ``invisible hand''.

\end{itemize}

for the crash is investigated. One may very well argue, as the author attempts
to do, that the two proposed mechanisms are two sides of the same thing, 
however, on quite different time scales.

\section{Statistics of LPPL Bubbles}

In a series of papers, the present author with D. Sornette and co-workers
\cite{JS2000,SJ2001,JS2002,SJB,SJ1997,JS1999,JSL1999,JLS2000} 
have shown that on the FX, Gold and major stock markets, crashes are often 
preceded by precursory characteristics quantified by a log-periodic power law, 
specifically
\be \label{lpeq}
p\lp t\rp = A+B\lp t_c - t\rp^z +C\lp t_c - t\rp^z \cos\lp  \omega 
\ln\lp t_c - t\rp - \phi \rp
\ee
The results on the US stock markets have been confirmed by several independent 
groups \cite{feigen,vandewalle,wolfgangpaul}. Eq. (\ref{lpeq}) has its origin in 
a Landau-expansion type of the argument and the underlying {\it Scaling Ansatz} 
is simply
\be \label{ansatz}
\frac{d F\lp x\rp}{d\ln x}=\alpha F\lp x\rp +\beta  F^2\lp x\rp \ldots
\ee
which to first order leads to eq. (\ref{lpeq}) with an arbitrary choice of
periodic function. The concept that only relative changes are important has a 
solid foundation in finance, but a detailed and rigorous derivation or 
justification of (\ref{ansatz}) has yet to be achieved. Instead,
the predictions that comes from applying this Ansatz, specifically those 
related to $\alpha = z + i \omega$, to the data have been compared using different 
markets and time periods. Specifically, assuming a Gaussian distribution it 
has been established empirically that 
\be \label{zomega}
\omega \approx 6.36 \pm 1.56 \hspace{10mm} z \approx 0.33 \pm 0.18
\ee
for over thirty crashes on major financial markets \cite{baliproc,JS2002}. 
Hence, (\ref{lpeq}) combined with (\ref{zomega}) is used as a definition of a
LPPL bubble. Together with the concepts of ``shocks'', as defined above, we are 
lead to a consistent and coherent picture when combined with a ranking of 
$\epsilon$-drawdowns.

\section{Statistics of Drawdowns}

As mentioned, an increasing amount of evidence that the largest negative market
moves belongs to a different population than the smaller has accumulated 
\cite{outl1,JS2000,outl2,SJ2001,baliproc,JS2002,Lillo}. Specifically, it was 
found that the cumulative distributions of drawdowns on the worlds major 
financial markets, {\it e.g.}, the U.S. stock markets, the Hong-Kong stock 
market, the currency exchange market (FX) and the Gold market are well 
parameterised by a stretched exponential
\be \label{stretched}
N(x) = Ae^{-bx^z} 
\ee
{\it except} for the $1\%$ (or less) largest drawdowns. In general, it was
found that the exponent $z \approx 0.7-0.9$ \cite{outl2}. It is worth noting 
that only the  distributions for the DJIA, the US\$/DM exchange rate and the 
Gold price exhibits ``potential'' outliers for the complement 
drawup distribution, whereas (all?) other markets shows a strong asymmetry 
between the tails of the drawdown and drawup distributions the latter having 
no outliers. This asymmetry has also been confirmed for smaller losses/gains
\cite{hori}. The range of the exponent for the drawup distributions is also
generally higher with $z \approx 0.9-1.1$ except for the FX and Gold markets 
\cite{outl2}.

In previous analysis and identification outliers on the major financial
markets drawdowns (drawups) were either simply defined as a continuous decrease 
(increase) in the closing value of the price \cite{outl2} or a simple filter of
$1\%$ and $2\%$ were used \cite{outl1,baliproc}. Hence, in the first case a drawdown 
(drawup) was terminated either by any increase (decrease) in the price no matter 
how small. In the second case the filter was arbitrarily chosen independently 
of the volatility of the data. Here an $\epsilon$-drawdown is defined as 
follows. A local maximum in the price is identified. Then a continuation of 
the downward trend ignoring movements in the reverse direction smaller than 
$\epsilon = a\sigma$ is identified. Here, sigma is the historical volatility 
of the index calculated from the data and $a=1$ for the DJIA and  $a=1/2$ for 
the SP500 and NASDAQ.

\section{Correspondence between Crashes, Bubbles and Shocks}

In figures 1-3 we see the cumulative $\epsilon$-drawdown distributions of the 
DJIA, SP500 and Nasdaq. The thresholds used were a relative threshold of 
$\epsilon = a\sigma$, where $\sigma$ is the historical volatility and $a$ is 
a number between 0 and 2, see figure legends. We see that the fits with eq. 
(\ref{stretched}) for all three index fully captures the distributions 
{\it except} for a few cases which is referred to as ``crashes''. In tables
\ref{outldj}-\ref{outlnas} we see the ranking of the $\epsilon$-drawdowns 
found in the three historical indices.

If we only note the dates of the four (DJIA), five (SP500) and six (NASDAQ) 
largest $\epsilon$-drawdowns in the tables, {\it i.e.},  $\epsilon \neq 0,2$, 
we get 
\begin{itemize}
\item DJIA: 1914, 1929, 1940 and 1987
\item SP500: 1946, 1962, 1987$\star$, 1998, 2000
\item Nasdaq: 1978, 1987$\star$, 1998$\star$, 2000
\end{itemize}
Here $\star$ denotes two $\epsilon$-drawdown linked to the same event. Except 
for the crashes related to the outbreak of WWI in 1914 and WWII in 1940 quite 
remarkably all of these outliers have log-periodic power law precursors 
well-described by eq. (\ref{lpeq}) and (\ref{zomega}). In addition, {\it all} 
have previously been published 
\cite{SJ2001,JS2002}. For the pure drawdowns, {\it i.e.},  $\epsilon = 0$,
the ranking order is permutated and for the DJIA the 1940 event is replaced by 
a 1933.55 event, for the SP500 the 1978 event is replaced by a 1974.72 event 
and for the NASDAQ the 1978 event is replaced by a 1990.62 event. These three
``new'' events coincide with the political maneuvering of president F. D. 
Roosevelt\footnote{Roosevelt New Deal policy included the passing of the 
Securities Acts of 1933 and 1934 as well as going of the Gold standard in 1933.} 
upsetting the financial markets, the political turmoil caused by the resignation 
and the controversial pardoning of president R. Nixon on August 8th and September 
8th 1974 and an $\epsilon$-drawdown of $17.8\%$ in the Japanese Nikkei index,
respectively, and should thus be classified as a shocks.

The results presented here means that the {\it joint evidence} from the 
ranking of $\epsilon$-drawdowns in the DJIA, the SP500 and the Nasdaq 
identifies all crashes with log-periodic power law precursors found in the 
US stock market {\it except} the crash after the LPPL bubble of 1937. Naturally, 
the optimal threshold (according to some specific definition) used in the outlier 
identification process is related to that particular index volatility. However, 
the volatility is again nothing but a measure of the two-point correlations 
present in the index, which we have proved to be an insufficient measure when 
dealing with extreme market events. This presumably explains this miss in the
1-to-1 correspondence between crashes on one hand and shocks and LPPL bubbles 
on the other.

\section{Conclusion}

The analysis presented here have strengthen the evidence for outliers in
the financial markets and that the concept can be used as a objective and
quantitative definition of a market crash. Furthermore, we have shown that
the existence of outliers in the drawdown distribution is primarily related 
to the existence of log-periodic power law bubbles prior to the occurrence of 
these outliers or crashes. In fact, of the 12 largest drawdowns identified as 
outliers only 3 did not have prior log-periodic power law bubble and these
3 outliers could be linked to a specific major historical event. In complement,
only 1 (1937) previously identified log-periodic power law bubble was not 
identified as an outlier.

the analysis presented suggest a one-to-one correspondence 
between market crashes defined as top-rank filtered drawdowns on one hand 
and surprising news and LPPL bubbles on the other. We attribute this 
correspondence to the Efficient Market Hypothesis effective on to quite 
different time scales depending on whether the market instability the crash
represent is internally or externally generated.

Further work is needed to clarify the role of different coarse-graining 
methods as well as to arrive at a quantitative choice for $\epsilon$ based
on the data. Naturally, the choice of $\epsilon$ should not only be determined
by the volatility but should also depend further on the type of market.

\newpage

Cited papers by the author are available from 
http://www.get2net.dk/kgs/pub.html.

\newpage

\begin{table}[]
\begin{center}
\begin{tabular}{|c|c|c|c|c|c|c|c|c|c|}\hline
$\epsilon$ & Date     & Size       & Duration &  Class &$\epsilon$ & Date     & Size     & Duration  &  Class\\ \hline
$0$        & 1987.786$\star$ & $30.7\%$   & 4 days  & Bubble & $\sigma$ & 1914.374 & $32.7\%$   & 64 days & Shock  \\ \hline
$0$        & 1914.579 & $28.8\%$   & 2 days    & Shock & $\sigma$ & 1987.786$\star$ & $30.7\%$   & 4 days  & Bubble   \\ \hline
$0$        & 1929.818$\star$ & $23.6\%$   & 3 days   &  Bubble & $\sigma$ & 1929.810$\star$ & $29.5\%$   & 6 days & Bubble\\ \hline
$0$        & 1933.549 & $18.6\%$   & 4 days   & Depression & $\sigma$ & 1940.261 & $23.7\%$   & 44days  & Shock  \\ \hline
\end{tabular}
\caption{\label{outldj}  List and properties of the highest ranked $\epsilon$-drawdowns 
for $\epsilon=0$ (left part) and $\epsilon=\sigma$ (right part) of the DJIA from 01/01 1900 
to 17/07 2000. The outliers are ranked by decreasing amplitudes. The term ``shock'' refers 
to an outlier which has been triggered by an event exogenous to the market. The term ``bubble'' 
embodies the idea that the corresponding $\epsilon$-drawdown  corresponds to a crash ending 
a speculative LPPL bubble.}
%\end{center}
%\end{table} 
\vspace{15mm}
%\begin{table}[]
%\begin{center}
\begin{tabular}{|c|c|c|c|c|c|c|c|c|c|}\hline
$\epsilon$ & Date             & Size       & Duration & Class & $\epsilon$  & Date             & Size       & Duration& Class\\\hline
$0$        & 1987.784$\star$ & $28.5\%$   & 4 days & Bubble  &  $\sigma/2$ & 1987.784$\star$ & $28.5\%$   & 4 days  & Bubble\\\hline
$0$        & 1962.370$\star$ & $13.7\%$   & 9 days & Bubble  &  $\sigma/2$ & 1946.636$\star$ & $16.2\%$   & 9 days  & Bubble\\\hline
$0$        & 1998.649$\star$ & $12.4\%$   & 4 days & Bubble  &  $\sigma/2$ & 1962.370$\star$ & $13.7\%$   & 9 days  & Bubble\\\hline
$0$        & 1987.805$\star$ & $11.8\%$   & 3 days & Bubble  &  $\sigma/2$ & 1998.649$\star$ & $12.4\%$   & 4 days  & Bubble\\\hline
$0$        & 1974.721         & $11.2\%$   & 9 days & Shock   &  $\sigma/2$ & 1987.805$\star$ & $11.9\%$   & 3 days  & Bubble\\\hline
\end{tabular}
\caption{\label{outlsp} Same as table \ref{outldj} but for the SP500 stock market index from 29/11 1940 to 17/07 2000.}
%\end{center}
%\end{table} 
\vspace{15mm}
%\begin{table}[]
%\begin{center}
\begin{tabular}{|c|c|c|c|c|c|c|c|c|c|}\hline
$\epsilon$ & Date     & Size            & Duration & Class & $\epsilon$ & Date             & Size       & Duration& Class  \\ \hline
$0$        & 2000.268$\star$ & $25.3\%$& 5 days   & Bubble& $\sigma/2$ & 1987.762$\dagger$ & $27.7\%$   & 11 days & Bubble \\ \hline
$0$        & 1987.784 $\dagger$ & $24.6\%$& 5 days   & Bubble& $\sigma/2$ & 2000.268$\star$ & $25.3\%$   & 5 days  & Bubble \\\hline
$0$        & 1987.805 $\dagger$ & $17.0\%$        & 5 days   & Bubble& $\sigma/2$ & 1998.630$\dagger$ & $19.2\%$   & 9 days  & Bubble \\\hline
$0$        & 1998.649$\dagger$ & $16.6\%$& 4 days   & Bubble& $\sigma/2$ & 1998.724$\dagger$ & $18.6\%$   & 9 days  & Bubble \\\hline
$0$        & 2000.374$\star$ & $14.9\%$& 5 days   & Bubble& $\sigma/2$ & 1987.806$\dagger$ & $17\%$     & 5 days  & Bubble \\\hline
$0$        & 1990.622 & $12.5\%$        & 6 days   & Shock & $\sigma/2$ & 1978.753$\dagger$& $16.6\%$   & 21 days & Bubble \\\hline
\end{tabular}
\caption{\label{outlnas} Same as table \ref{outldj} but for the NASDAQ stock market index from 05/02 1971 to 17/07 2000.}  
\end{center}
\end{table}

\newpage

\begin{figure}
\begin{center}
\vspace{-10mm}
\parbox[]{13.5cm}{
\epsfig{file=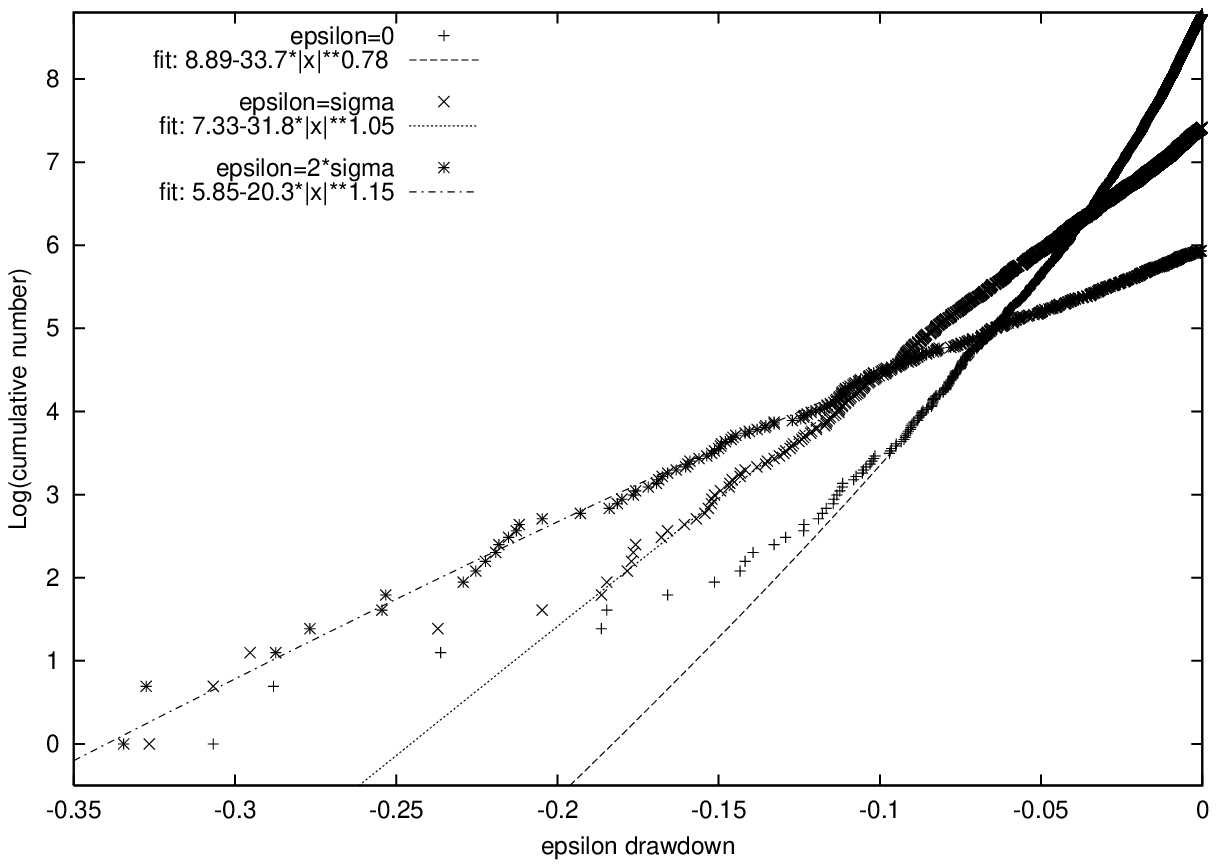,height=7.5cm,width=13.5cm}
\caption{$\epsilon$-drawdown distribution for DJIA from 01/01 1900 to 17/07 2000.} }
\parbox[]{13.5cm}{
\vspace{5mm}
\epsfig{file=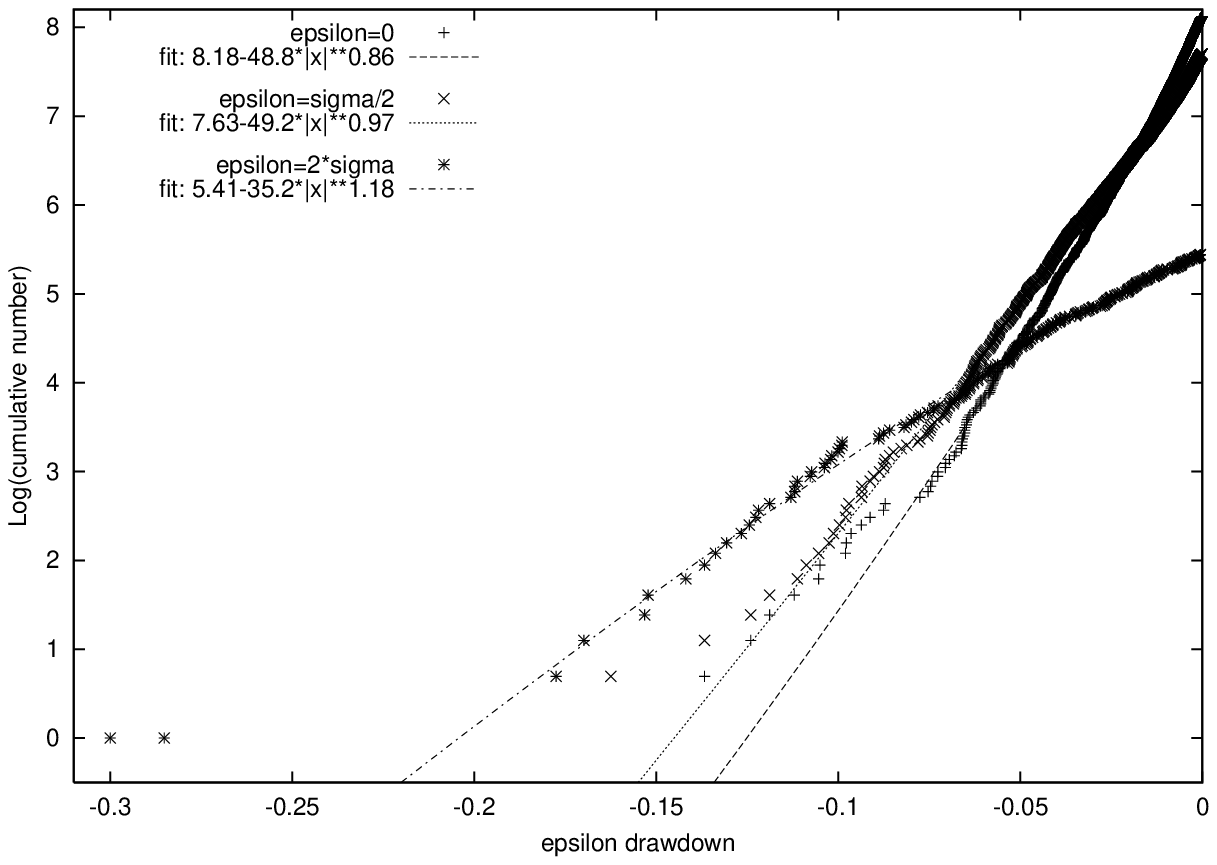,height=7.5cm,width=13.5cm}
\caption{$\epsilon$-drawdown distribution for SP500 from 29/11 1940 to 17/07 2000.} }
\parbox[]{13.5cm}{
\vspace{5mm}
\epsfig{file=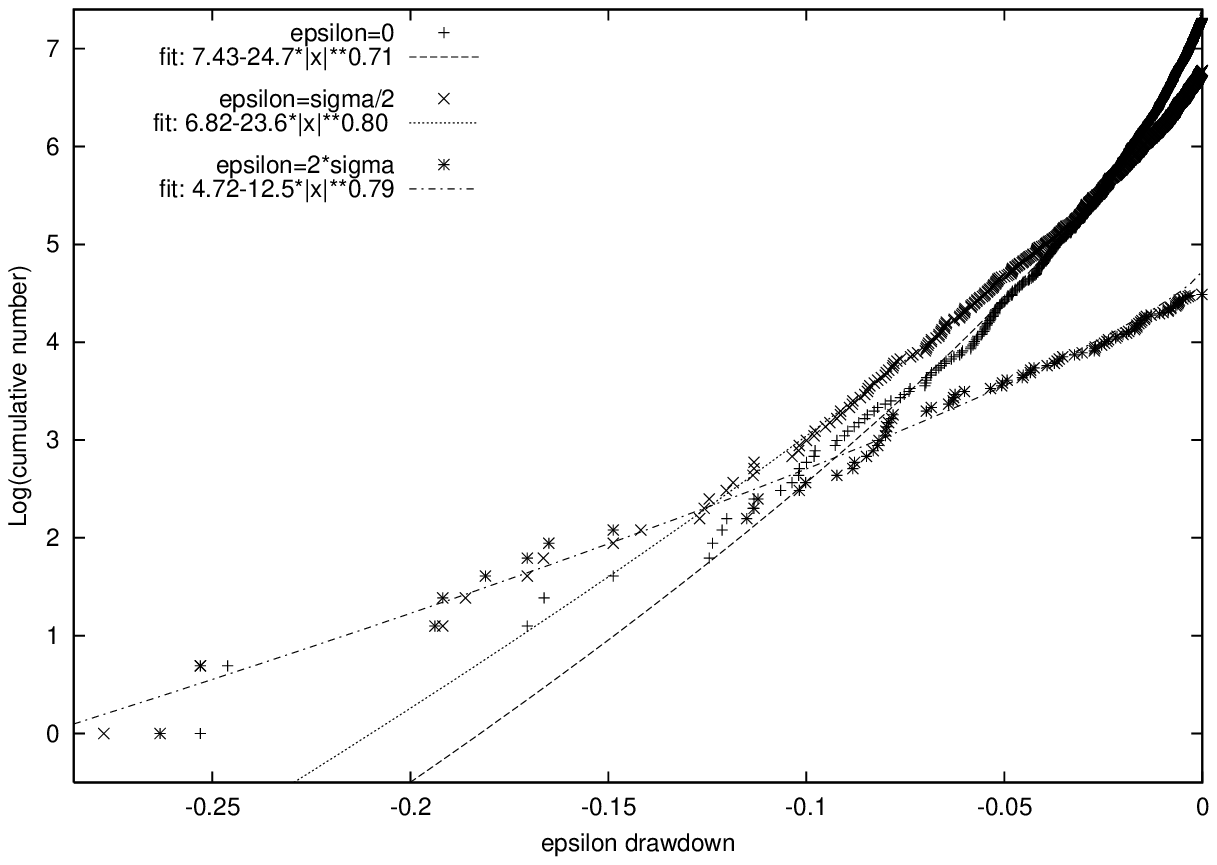,height=7.5cm,width=13.5cm}
\caption{$\epsilon$-drawdown distribution for NASDAQ from 05/02 1971 to 17/07 2000.} }
\end{center}
\end{figure}

\end{document}